\documentstyle{article}
\newtheorem{theorem}{Theorem}[section]

\newenvironment{prooof}{\begin{description}
                   \item[{\small {\bf Proof:}}] \small}{\hfill {\bf Q.E.D.}
                                                          \medskip
                                                       \end{description}}

\newtheorem{defi}{Definition}[section]
\newtheorem{prop}{Proposition}[section]
\newtheorem{lemma}{Lemma}[section]
\newtheorem{rem}{Remark}[section]

\newcommand{\bdef}{\begin{defi}}
\newcommand{\ede}{\end{defi}}
\newcommand{\bsat}{\begin{theorem}}
\newcommand{\esat}{\end{theorem}}
\newcommand{\bprop}{\begin{prop}}
\newcommand{\eprop}{\end{prop}}
\newcommand{\blem}{\begin{lemma}}
\newcommand{\elem}{\end{lemma}}
\newcommand{\brem}{\begin{rem}}
\newcommand{\erem}{\end{rem}}
\newcommand{\bbew}{\begin{prooof}}
\newcommand{\ebew}{\end{prooof}}
\newcommand{\be}{\begin{equation}}
\newcommand{\ee}{\end{equation}}

\newcommand{\ra}{\rightarrow}

\newcommand{\f}{\frac}
\newcommand{\p}{\partial}

\newcommand{\Real}{\mbox{I \hspace{-0.82em} R}}

\newcommand{\df}{\stackrel{\rm def}{=}}

\begin{document}
\title{The Dynamics of Relativistic Membranes\\
         II: Nonlinear Waves and Covariantly Reduced Membrane
             Equations \vspace{1cm}}
\author{{\bf Martin Bordemann}\\Fakult\"at f\"ur Physik\\Universit\"at
                             Freiburg\\
          Hermann-Herder-Str. 3\\79104 Freiburg i.~Br., F.~R.~G\\
          e-mail: mbor at ibm.ruf.uni-freiburg.de
        \and
        {\bf Jens Hoppe}\\Institut f\"ur Theoretische Physik\\Universit\"at
           Karlsruhe\\P.~O.~Box 6980\\76137 Karlsruhe, F.~R.~G.\\
           e-mail: be10 at dkauni2.bitnet
       }
\date{FR-THEP-93-19 \\ KA-THEP-5-93 \\ August 1993}
\maketitle
\begin{abstract}

By explicitly eliminating all gauge degrees of freedom in the
$3+1$-gauge description of a classical relativistic (open) membrane
moving in $\Real^3$ we derive a $2+1$-dimensional nonlinear wave equation
of Born-Infeld type for the graph $z(t,x,y)$ which is invariant under the
Poincar\'e group in four dimensions.
Alternatively, we determine the world-volume of a membrane in a
covariant way by the zeroes of a scalar field $u(t,x,y,z)$ obeying a
homogeneous Poincar\'e-invariant nonlinear wave-equation. This
approach also gives a simple derivation of the nonlinear gas dynamic
equation obtained in the light-cone gauge.
\end{abstract}
\vfill
\newpage

{\bf 1.} Generalizing the Nambu-Goto action to describe higher-dimensional
relativistically invariant minimal surfaces in
$D$-dimensional Minkowski space, the classical equations of motion
constitute a set of $D$ coupled, highly nonlinear, partial differential
equations for $D$ functions $x^{\mu}(\varphi^0,\ldots,\varphi^M)$ of $M+1$
variables that parametrize the minimal surface. The solution of these
equations is hindered by the large, non-explicit, redundancy resulting
from the gauge degrees of freedom that reflect the invariance under the
diffeomorphism group reparametrizing $(\varphi^0,\ldots,\varphi^M)$.
Previous
attempts to solve the equations have, so far, mainly been confined to
the light cone description, due to the fact that one
of the unknown functions can then be eliminated. In a recent paper
\cite{BH93} the constraints arising from this elimination were solved
for the case
$D=4$, $M=2$, and the theory was shown to be equivalent
to a 2+1-dimensional inviscid irrotational isentropic gas whose
pressure is inversely proportional to minus the gas density.

In this letter, we would like to consider the following,
slightly more geometrical situation: choosing $\varphi^0=x^0$, the time
dependent shape of the $M$-dimensional surface, as seen by an
observer with time $t=x^0$, is given by
$x^i(t,\varphi^1,\ldots,\varphi^M)$, $i=1,\ldots,D-1$. A further gauge
choice (noted already in \cite{Hop82}) allows one to assume that the
$D-1$-dimensional velocity vector
$\dot{x}^i(t,\vec{\varphi})$ is normal to the surface at
$(t,\vec{\varphi})$ {\em and} that the purely spatial part of the
induced metric equals $1-\dot{\vec{x}}^2$ for all $t$. We shall refer to
this kind of gauge (cf. \cite{Hop82}) as the ``orthonormal 3+1-gauge''.

In the case $D=4,M=2$ we arrive at a first order vector equation for the
spatial part $\vec{x}$ which in particular implies that $\dot{\vec{x}}$
is proportional to
the surface normal in $\Real^3$ (if this dynamical proportionality
factor was equal to
$|\p_1\vec{x}\times\p_2\vec{x}|$ this equation would coincide with
the $SU(\infty)$-Nahm-equation which was shown to be linearizable
by Ward (cf. \cite{War90})). The vector equation may now be reduced
by a hodograph technique similar to the one we applied to the
equations of motion in the light cone gauge
(\cite{BH93}); finally we deduce a second order equation for $z=x^3$
as a function of $t$ and $(x^1,x^2)$ which can be derived from the
Born-Infeld type Lagrangean $\sqrt{1-\p^{\alpha}z\p_{\alpha}z}$.
Despite its appearance as a covariant $2+1$-dimensional equation
it actually does have the full $3+1$-dimensional Poincar\'e invariance.

In a second part of this letter we show that the equation for $z$
can be derived in a more direct way from a manifestly covariant nonlinear
$3+1$-dimensional wave equation
for a scalar function $u(t,x^1,x^2,x^3)$ whose hypersurfaces $u=const$
are all minimal hypersurfaces, i.~e. solutions of the membrane
equation.
The Lagrangean for $u$ is simply given by the homogeneous
functional $\sqrt{(\p u)^2}$ times the space-time volume. The derivation
is valid for a curved background.

Finally, we show that the second order equation
for the velocity potential $p$ coming from the reduction of the membrane
equation in the light cone gauge (see \cite{BH93}) can also be derived
from the above-mentioned covariant equation for $u$. The
Lagrangean for $p$ is given by $\sqrt{\dot{p}+\f{1}{2}(\vec{\nabla}p)^2}$.

\vspace{0.5cm}

{\bf 2.} Let $\Sigma$ be an $M$-dimensional manifold with co-ordinates
$(\varphi^1,\ldots,\varphi^M)$ and $({\cal M},\eta)$ be a $D$-dimensional
(possibly curved) Lorentz manifold,
and the action for the world-volume
$x=(x^{\mu}):\Real\times\Sigma\ra{\cal M}$ of the surface $\Sigma$
moving in $\cal M$ be given by the $M+1$-dimensional volume swept out
in space-time:
\be S[x]~=~\int\!d\varphi^0 d^M\varphi~\sqrt{G} \hspace{1cm}; \ee
$G$ is $(-1)^M$ times the determinant of the induced world-volume
metric
$G_{\alpha\beta}$ (which we assume to be nondegenerate throughout this
paper):
\be G_{\alpha\beta}~\df~
         x^{\mu},_{\alpha}x^{\nu},_{\beta}\eta_{\mu\nu}(x)
         \hspace{1cm}. \ee
The index notation will always be
($\alpha,\beta,\gamma,\ldots=0,1,\ldots,M$ and
$\lambda,\mu,\nu,\ldots=0,1,\ldots,D-1$), and
as usual, a comma in front of an index denotes partial derivative with
respect to the corresponding variable.
The resulting field equations are
\be \frac{1}{\sqrt{G}}
     (\sqrt{G}G^{\alpha\beta}x^{\mu},_{\alpha}),_{\beta}
     ~+~G^{\alpha\beta}x^{\nu},_{\alpha}x^{\rho},_{\beta}
           \Gamma^{\mu}_{\nu\rho}(x)
            ~=~0 \hspace{1cm}, \label{field1} \ee
where $\Gamma^{\mu}_{\nu\rho}$ denote the Christoffel symbols of the
metric $\eta$ (which vanish if $\cal M$ is flat Minkowski space and
$x^{\mu}$ are the standard co-ordinates). The
following alternative
formulation of the field equations turns out to be useful later: Let
\be (\Pi_{\perp})^{\mu}_{\nu}(x)~\df~\delta^{\mu}_{\nu}-
        x^{\mu},_{\alpha}x^{\rho},_{\beta}G^{\alpha\beta}\eta_{\rho\nu}(x)
        \label{proj} \ee
be the orthogonal projection on the subspace orthogonal to the tangent
space of the world-volume in $\cal M$. Eqn (\ref{field1}) may then be
written as (cf. e.~g. \cite{Eis64}, p.~178):
\be (\Pi_{\perp})^{\mu}_{\nu}(x)
       \big( x^{\nu},_{\alpha\beta}
       +\Gamma^{\nu}_{\rho\sigma}(x)x^{\rho},_{\alpha}x^{\sigma},_{\beta}
       \big)G^{\alpha\beta}~=~0 \hspace{1cm}, \label{field2}  \ee
showing that only the equations along the orthogonal
components are truly dynamical (in particular, there would be
no such equation if $M+1=D$).
Both (\ref{field1}) and (\ref{field2}) are invariant under both arbitrary
reparametrizations $\Real\times\Sigma\ra\Real\times\Sigma$ of the
world-volume and isometries ${\cal M}\ra{\cal M}$.

\vspace{0.5cm}

{\bf 3.} Let us now reduce the dynamical equations (\ref{field1}) for a
two-dimensional membrane moving in four-dimensional Minkowski space
$\Real^{(1,3)}$ by first choosing $\varphi^0=x^0\df t$ (``3+1-gauge'').
Denoting by $\vec{x}$ the spatial part of the four-vector $x$
we get ($r,s=1,2$)
\be G_{\alpha\beta}=\left( \begin{array}{cc}
                     1-\dot{\vec{x}}^2 & -\dot{\vec{x}}\cdot\vec{x},_r \\
         -\dot{\vec{x}}\cdot\vec{x},_r & -\vec{x},_r\cdot\vec{x},_s
                           \end{array} \right) \ee
and
\be G~=~(\vec{x},_1\times\vec{x},_2)^2~-~
             (\dot{\vec{x}}\cdot(\vec{x},_1\times\vec{x},_2))^2
             \hspace{1cm}. \ee
One then observes that the spatial part of the field equations
(\ref{field1}) ($\mu=i=1,2,3$) is orthogonal to both tangent vectors
$\vec{x},_1$ and $\vec{x},_2$, and that its component normal to the
embedded surface
coincides with the temporal part of eqn (\ref{field1}) ($\mu=0$),
which in the gauge (\cite{Hop82}, p.~149)
\be \dot{\vec{x}}\cdot\vec{x},_r~=~0~~~~~~{\rm for}~~~r=1,2
                        \label{tangauge} \ee
reads
\be \f{\p}{\p t}\left(\sqrt{\f{g}{1-\dot{\vec{x}}^2}}~\right)~=~0
                        \label{resteqn} \hspace{1cm}, \ee
where $g$ denotes the determinant of the spatial metric $g_{rs}\df
\vec{x},_r\cdot\vec{x},_s$,
\be g~=~(\vec{x},_1\times\vec{x},_2)^2  \label{eqng} \hspace{1cm}. \ee
As the gauge
choice (\ref{tangauge}) still leaves the freedom of performing
time-independent reparametrisations $\Sigma\ra\Sigma$, one may choose
$g$ (at a particular time) to be actually {\em equal} to
$1-\dot{\vec{x}}^2$ (\cite{Hop82}). Equation
(\ref{resteqn}) then
implies that this choice is preserved in time, thus eliminating the
arbitrary time-independent function $\Sigma\ra\Real$ allowed by
(\ref{resteqn}). Combining this choice and the gauge (\ref{tangauge})
to a single first order vector equation one gets
\be \dot{\vec{x}}~=~\sqrt{\f{1}{(\vec{x},_1\times\vec{x},_2)^2}-1}
                          ~ (\vec{x},_1\times\vec{x},_2)
                           \hspace{1cm}. \label{veceqn} \ee
Written in the form
\be \dot{x_i}=\f{\gamma}{2}\epsilon_{ijk}\{x^j,x^k\} \label{Nahm} \ee
(where $\gamma\df\sqrt{\f{1}{g}-1}$ and $\{~,~\}$ is the Poisson
bracket with respect to the variables $\varphi^1$ and $\varphi^2$ on
$\Sigma$) this equation would be equal to the
$su(\infty)$-Nahm equation (known to be integrable \cite{War90}) if the
prefactor $\f{\gamma}{2}$ was 1.

Equations (\ref{Nahm}) are still invariant under area-preserving
reparametrizations $\Sigma \ra\Sigma$ (i.~e. reparametrizations whose
Jacobi determinant
is equal to 1). In order to remove this residual degree of gauge freedom
let us give a description of the two-dimensional surface at a given
time in terms of the graph $(x^1,x^2,z(t,x^1,x^2))$, rather than in terms
of the three Cartesian co-ordinates as functions of the ``fictitious''
parameters $(\varphi_1,\varphi_2)$. We therefore perform the following
change of variables (which had already turned out to be useful in the
light-cone description (cf. \cite{BH93})):
\be (\varphi^0=t,\varphi^1,\varphi^2)
             \mapsto(t=x^0,x^1(t,\vec{\varphi}),x^2(t,\vec{\varphi})) \ee
where we have set $\vec{\varphi}=(\varphi^1,\varphi^2)$. Observing that
the Jacobi determinant $J$ of this transformation is equal to the
Poisson bracket $\{x^1,x^2\}$ and that (due to eqn (\ref{Nahm}), $i=1,2$)
the ``old'' time derivative $\f{\p}{\p t}$ becomes
$\f{\p}{\p x^0}-\gamma J \vec{\nabla}z\cdot\vec{\nabla}$,
the $i=3$ part of eqn
(\ref{Nahm}) becomes (where $\dot{~}$ from now on denotes the ``new''
time derivative and $\vec{\nabla}$ stands for the gradient with respect to
$x^1$ and $x^2$):
\be \dot{z}=\gamma J(1+(\vec{\nabla}z)^2)  \hspace{1cm}, \label{zeqn}  \ee
while the two equations corresponding to $i=1,2$  in eqn (\ref{Nahm})
imply
\be \dot{J}=-J^2\vec{\nabla}\cdot(\gamma \vec{\nabla}z)
                                 \hspace{1cm}; \label{Jeqn}  \ee
in the new variables $\gamma $ can be expressed as
\be \gamma = \sqrt{J^{-2}(1+(\vec{\nabla}z)^2)^{-1}-1}
                                 \hspace{1cm}. \ee
Rewriting eqn (\ref{Jeqn}) in a form involving only $z$, its derivatives,
and $\gamma J$ (for which we can use eqn (\ref{zeqn})) we get the following
second order equation for $z$:
\be \ddot{z}-\vec{\nabla}^2z~=~
    (\vec{\nabla}z)^2(-\ddot{z}+\vec{\nabla}^2z)
       -\f{1}{2}\vec{\nabla}z\cdot\vec{\nabla}(\vec{\nabla}z)^2
       +\vec{\nabla}\dot{z}^2\cdot\vec{\nabla}z
       -\dot{z}^2\vec{\nabla}^2z  \hspace{1cm}. \label{Bornin} \ee
This is nothing but the Euler-Lagrange equation for a $2+1$-dimensional
scalar field theory described by a Lagrangean density
\be {\cal L}~\df~-\sqrt{1-\dot{z}^2+(\vec{\nabla}z)^2}
            ~=~-\sqrt{1-z,_{\alpha}z^{,\alpha}} \hspace{1cm};
                                       \label{lagbi} \ee
here the $\alpha$ indices range over 3-dimensional Minkowski space
$\Real^{(1,2)}$.
The corresponding Hamiltonian is given by
\be H=\int\!d^2x~\sqrt{1+\pi^2}\sqrt{1+(\vec{\nabla}z)^2}
                                                     \label{hambi} \ee
where $\pi = \delta {\cal L}/\delta \dot{z}$ is the momentum conjugate
to the field $z$. Note that an expansion of the square root in
eqn (\ref{hambi}) will give a free field theory in lowest nonconstant
order. Long ago, such nonlinear field theories had been investigated by
Born and Infeld (\cite{BI}) and Heisenberg (\cite{Hei}). Later on, in
\cite{BC}, the integrability of the $1+1$-dimensional theory ($z(t,x)$)
was noted (which is clear by the hodograph trick, compare \cite{Bog89}
or \cite{Whi74}, p. 617),
and a quantum theory was developed (cf. \cite{BC}). While the
$SO(1,2)$-invariance of the Lagrangean (\ref{lagbi}) and
the field equations (\ref{Bornin}) is obvious,
\be (1-z^{,\alpha }z,_{\alpha })z^{,\beta},_{\beta}
       +z^{,\alpha}z^{,\beta}z,_{\alpha \beta}~=~0
                              \label{Bornincov}   \hspace{1cm}, \ee
it was apparently
{\em not} noticed that the theory has a much wider higher dimensional
invariance group. In order to see the invariance under the
Lorentz group $SO(1,3)$ one remembers that $z$ is the 3-component of a
four-vector, $x^{\mu}=(t,x^1,x^2,z(t,x^1,x^2))^T$, and transforms this
four-vector in the usual way by a Lorentz transformation
$\Lambda^{\mu}_{\nu}$, i.~e.
\be \tilde{x}^{\mu}=(\tilde{t},\tilde{x}^1,\tilde{x}^2,
                      \tilde{z}(\tilde{t},\tilde{x}^1,\tilde{x}^2))^T
                   =\Lambda^{\mu}_{\nu}x^{\nu}  \hspace{1cm},
                                 \label{tranglo} \ee
implicitly determining the function $\tilde{z}$.
The infinitesimal action of the Lie algebra of the Lorentz
group (obtained by setting $\Lambda=e^{\epsilon K}$ and
differentiating eqn (\ref{tranglo}) w.~r.~t. $\epsilon$ at $\epsilon=0$)
reads:
\be (\delta_Kz)(t,x^1,x^2)=
       -\p_{\alpha}z(t,x^1,x^2)
         (K^{\alpha}_{\beta}x^{\beta}+K^{\alpha}_3z(t,x^1,x^2))
       +K^3_{\beta}x^{\beta}  \hspace{0.3cm}. \label{traninf} \ee

The infinitesimal boost in the $z$-direction, for instance,
results in
$(\delta_Kz)(x^{\alpha})=-(\p_tz)(x^{\alpha})z(x^{\alpha})+t$ whereas an
infinitesimal rotation in the $x^1-z$-plane gives
$(\delta_Kz)(x^{\alpha})=-(\p_1z)(x^{\alpha})z(x^{\alpha})-x^1$. In order to
see that the transformation (\ref{traninf}) is a symmetry of eqn
(\ref{Bornincov}) one linearizes the l.~h.~s. of (\ref{Bornincov})
getting a linear equation for $(\delta_Kz)(t,x^1,x^2)$ with coefficients
depending on $z$ which gives zero after inserting (\ref{traninf}) and
using the fact that $z$ solves (\ref{Bornincov}).

It is also not difficult to write down the corresponding conserved charges,
e.~g. (for the boost in the $z$-direction):
\be Q_{tz}~\df~ t\int\!d^2x~\pi~-~\int\!d^2x~z\sqrt{1+(\vec{\nabla}z)^2}
                                    \sqrt{1+\pi^2} \hspace{1cm}. \ee
Along these lines one can verify that
\be M^2~\df~H^2~-~(\int\!d^2x~\pi\vec{\nabla}z)^2~-~(\int\!d^2x~\pi)^2 \ee
is Poincar\'e-invariant.

\vspace{0.5cm}

{\bf 4.} We shall now deduce eqn (\ref{Bornincov}) from a single
four-dimensional scalar equation:

The crucial observation is that the world-volume of a two-dimensional
membrane moving through four-dimensional Minkowski space is a minimal
{\em hypersurface} (i.~e. of codimension 1) and that hypersurfaces
may alternatively be described by the zero set of a smooth function (an
idea
which is already mentioned, though not pursued, in Dirac's paper
\cite{Dir62}). Considering again the case of a curved Lorentz
manifold $({\cal M},\eta)$
of arbitrary dimension, suppose
that there is a smooth function $u:{\cal M}\ra\Real$ such that its zero set
$\{r\in {\cal M}|u(r)=0\}$ is a minimal hypersurface. Parametrizing this
hypersurface by $(\varphi^0,\ldots,\varphi^M)$ means that
\be u(x(\varphi))=0  \label{uzero} \hspace{1cm}, \ee
which by differentiating with respect to the parameters
$\varphi^{\alpha}$ implies:
\be u,_{\mu}(x)x^{\mu},_{\alpha}=0 \hspace{1cm}.  \label{duzero} \ee
Assuming that $\eta^{\mu\nu}(x)u,_{\mu}(x)u,_{\nu}(x)\neq 0$
we immediately get a formula for the projector (\ref{proj}) on
the one-dimensional space orthogonal to the surface:
\be (\Pi_{\perp})^{\mu}_{\nu}(x)=
          \delta^{\mu}_{\nu}-
        x^{\mu},_{\alpha}x^{\rho},_{\beta}G^{\alpha\beta}\eta_{\rho\nu}(x)
       =\f{\eta^{\mu\rho}(x)u,_{\rho}(x)u,_{\nu}(x)}
                   {\eta^{\kappa\lambda}(x)u,_{\kappa}(x)u,_{\lambda}(x)}
                              \label{uproj} \hspace{0.3cm}. \ee
Now we can reformulate the field equations in its projector form
(\ref{field2}):
\begin{eqnarray*}
 0 & = & u,_{\lambda}(x)(x^{\lambda},_{\alpha\beta}
                     +\Gamma^{\lambda}_{\mu\nu}(x)
                        x^{\mu},_{\alpha}x^{\nu},_{\beta})G^{\alpha\beta} \\
   & = & \big( (u,_{\lambda}(x)x^{\lambda},_{\alpha}),_{\beta}
            -u,_{\lambda\kappa}(x)x^{\lambda},_{\alpha}x^{\kappa},_{\beta}
            +u,_{\lambda}(x)\Gamma^{\lambda}_{\mu\nu}(x)
              x^{\mu},_{\alpha}x^{\nu},_{\beta} \big) G^{\alpha\beta}  \\
   & = & 0~-~u_{;\mu\nu}(x)x^{\mu},_{\alpha}x^{\nu},_{\beta}
                    G^{\alpha\beta}  \\
   & = & -u_{;\mu\nu}(x)(\eta^{\mu\nu}(x)-(\Pi_{\perp})^{\mu\nu}(x))
\end{eqnarray*}
where the semicolon ; denotes the covariant derivative with respect to the
metric $\eta$. Using eqn (\ref{uproj}) this finally results in
\be 0~=~(\p u)^2\Box u-u^{,\kappa}u^{,\lambda}u_{;\kappa \lambda}
               \label{field3} \ee
where $(\p u)^2$ stands for $\eta^{\mu \nu}u,_{\mu}u,_{\nu}$. A priori,
this equation
holds on the world-volume $u=0$ only .

However, it can be shown that locally one gets no restrictions when
eqn (\ref{field3}) is allowed to hold outside the world-volume: Firstly, we
easily see
that this equation is then invariant under the group of isometries
of $({\cal M},\eta)$. Moreover,
the above derivation shows that {\em every} hypersurface $u=const$ is a
minimal hypersurface if $(\p u)^2\neq 0$, because eqn (\ref{duzero}) is still
valid. Consequently, every solution of
(\ref{field3}) regarded on all of $\cal M$ results in a foliation of
$\cal M$ into minimal hypersurfaces.
In fact, such a foliation exists as a local extension of any given local
minimal hypersurface at least for homogeneous $\cal M$ (such as
e.~g. Minkowski space): suppose one is given a small piece of a minimal
hypersurface in $\cal M$. Then (after shrinking the piece if necessary)
there exists a Killing field of the metric $\eta$ which is
nowhere tangent to this surface patch. The translates by the
flow of the Killing
field of the surface patch will all be minimal hypersurfaces, hence a
small open neighbourhood of the initial surface patch is indeed foliated
into minimal hypersurfaces. Any (locally defined) function $u$ ``counting''
the leaves of this foliation
satisfies eqn (\ref{field3}). A rather elementary example is provided
by the world-volume $(t,x,y)\mapsto(t,x,y,0)$ of the open $\Real^2$-type
membrane. Clearly, this is the zero set of the linear function $u(x^{\mu})
=z$ which obviously satisfies the nonlinear wave equation (\ref{field3}).
But also the affine hyperplanes $u=z_0\neq0$ are minimal world-volumes
(being translates along the $z$-direction).

Note that eqn (\ref{field3}) is also invariant under redefinitions
$u\mapsto s(u)$ where $s$ is any smooth real-valued function of $\Real$.

As can easily be computed, eqn (\ref{field3}) can be derived from the
Lagrangean
\be  {\cal L}~\df~\sqrt{\det(\eta_{\mu\nu})\eta^{\kappa\lambda}
                                 u,_{\kappa}u,_{\lambda}}
                              \hspace{1cm}. \label{unilag} \ee
This Lagrangean is homogeneous of weight one and (for flat
$\cal M$) therefore falls into
the class of Lagrangeans considered by Fairlie et al (\cite{Fairetal}).
This means in particular that any solution of the field equation
(\ref{field3}) is automatically a solution of the 4-dimensional
universal field equation
\be u^{,\mu}(\p^2u)^{-1}_{\mu\nu}u^{,\nu}~=~0 \ee
whose general solution is known (again by a hodograph transformation, cf.
\cite{Fair93}).

In order to see how
equation (\ref{Bornincov}) can be derived from
eqn (\ref{field3}) consider again four-dimensional Minkowski space
$\Real^{(1,3)}$. Take
a solution $u$ of the above field equation (\ref{field3}) and consider
\be
  0 = u(t,x^1,x^2,z(t,x^1,x^2))  \hspace{1cm}, \label{zneu}
\ee
which implicitly defines the scalar function $z(t,x^1,x^2)$.
Differentiating this equation once with respect to $(t,x^1,x^2)$ one can
express the first derivatives of $z$ and in terms of first derivatives
of $u$: $z,_{\alpha}=-u,_{\alpha}/u,_3$, while
differentiating twice one can also express
second derivatives of $z$ by first and second derivatives of $u$.
After some calculation one finds that the above-defined function $z$
indeed satisfies the field equations (\ref{Bornincov})
as a consequence of eqn (\ref{field3}).

\vspace{0.5cm}

{\bf 5.} In \cite{BH93} we had derived equations of the following
isentropic irrotational inviscid gas-dynamic type using the light-cone gauge:
$\dot{q}+\vec{\nabla}\cdot(q\vec{\nabla}p) = 0$ and
$\dot{p}+\frac{1}{2}(\vec{\nabla}p)^2-F(q) = 0$
where $q$ and $p$ are functions of time and $M$ spatial variables and
$F$ is a monotonous function $\Real\ra\Real$ such that the pressure
$P$ depends via $P'(q)=-qF'(q)$ on the gas density $q$. In our case of
four-dimensional Minkowski space we had $M=2$ and
$F(q)=\frac{1}{2}q^{-2}$. Solving the second equation for $q$ and
inserting the result in the first one gets
a second order equation for the velocity potential $p$ alone
which is apparently called ``Steichen equation'' (cf. \cite{Zey91}, p.~45).
A little computation reveals that this equation can be
derived from the Lagrangean
${\cal L}~\df~
    h\big(\dot{p}+\frac{1}{2}(\vec{\nabla}p)^2\big)$ where the real-valued
function $h$ of one real variable is such that its derivative equals the
inverse function of $F$.
In particular, for an adiabatic relation like
$P(q)=aq^{\gamma}$ with $a$ nonzero and $0 \neq \gamma \neq 1$ one
would arrive at
${\cal L}=(\dot{p}+\frac{1}{2}(\vec{\nabla}p)^2)^{\frac{\gamma}{\gamma-1}}$
up to a constant prefactor. For the membrane one has $\gamma=-1$,
hence
\be {\cal L}~=~\sqrt{\dot{p}+\frac{1}{2}(\vec{\nabla}p)^2}
     \hspace{1cm}. \label{lagmemlc}  \ee
This leads to the following equation for the membrane:
\be \ddot{p}~+~2\vec{\nabla}p\cdot\vec{\nabla}\dot{p}~+~
  \vec{\nabla}p\cdot\big( (\vec{\nabla}p\cdot\vec{\nabla})
   \vec{\nabla}p\big)~-~
    2(\dot{p}+\frac{1}{2}(\vec{\nabla}p)^2)\vec{\nabla}^2 p
    ~~=~~0 \hspace{0.5cm}. \label{Steichenmem}  \ee

This equation can also be derived from the covariant
scalar field equation (\ref{field3}): set
\be
  0 = u(t+\f{p(t,x^1,x^2)}{2},x^1,x^2,t-\f{p(t,x^1,x^2)}{2})
                            \hspace{1cm}. \label{pneu}  \ee
Again the first and second derivatives of $p$ can be expressed in terms of
first and second
derivatives of $u$ by differentiating this equation with respect to
$(t,x^1,x^2)$ getting for instance
$\dot{p}=-2(\p_0u+\p_3u)/(\p_0u-\p_3u)$, $p,_i=-2u,_i/(\p_0u-\p_3u)$.

Note that -- as in the case for the function $z$ -- one can deduce the
Lorentz symmetry of eqn (\ref{Steichenmem}) (cp. \cite{BH93}, eqs
(44)--(47) for the Hamiltonian treatment) by transforming the four-vector
$(t+\f{p(t,x^1,x^2)}{2},x^1,x^2,t-\f{p(t,x^1,x^2)}{2})^T$ by an arbitrary
Lorentz transformation $\Lambda^{\mu}_{\nu}$. The Poincar\'e invariance,
and the scaling symmetries of eqn (\ref{Steichenmem}), has also been
obtained by computer algebraic calculations (cf. \cite{Cla93}).

\vspace{0.4cm}

\noindent {\bf\Large Acknowledgment}

\vspace{0.2cm}

\noindent The authors would like to thank P.~Clarkson, J.~Eggers, T.~Filk,
B.~Fuchssteiner, D.~Giulini, B.~G.~Konopelchenko, and M.~Wadati for
valuable discussions.

\end{document}